\documentclass[prd,superscriptaddress,amsfonts,amssymb,amsmath,showpacs,onecolumn]{revtex4-2}
\usepackage{bm}
\usepackage{amsfonts}
\usepackage{latexsym}
\usepackage[utf8]{inputenc}
\usepackage{graphicx}
\usepackage{amsmath}
\usepackage{palatino}
\usepackage{mathpazo}
\usepackage{textcomp}
\usepackage{float}
\usepackage{booktabs}
\usepackage{dcolumn}
\usepackage{ragged2e}
\usepackage{hyperref}
\hypersetup{colorlinks,citecolor=blue}
\hypersetup{colorlinks=true,linkcolor=blue,filecolor=magenta,urlcolor=blue}
\usepackage{amsmath}
\usepackage{subfigure}
\usepackage[]{natbib}
\usepackage{xcolor}
\usepackage{orcidlink}
\usepackage{epsfig}
\usepackage{caption}
\usepackage[toc]{appendix}
\usepackage{commath}
\usepackage{cancel}
\usepackage{csquotes}
\usepackage{placeins}

\usepackage{multirow}

\def\jnl@style{\it}
\def\aaref@jnl#1{{\jnl@style#1}}

\def\aaref@jnl#1{{\jnl@style#1}}

\def\aj{\aaref@jnl{AJ}}                   
\def\apj{\aaref@jnl{ApJ}}                 
\def\apjl{\aaref@jnl{ApJ}}                
\def\apjs{\aaref@jnl{ApJS}}               
\def\apss{\aaref@jnl{Ap\&SS}}             
\def\aap{\aaref@jnl{A\&A}}                
\def\aapr{\aaref@jnl{A\&A~Rev.}}          
\def\aaps{\aaref@jnl{A\&AS}}              
\def\mnras{\aaref@jnl{Mon.~Not.~Roy.~Astron.~Soc.}}             
\def\prd{\aaref@jnl{Phys.~Rev.~D}}        
\def\prc{\aaref@jnl{Phys.~Rev.~C}}  
\def\prl{\aaref@jnl{Phys.~Rev.~Lett.}}    
\def\qjras{\aaref@jnl{QJRAS}}             
\def\skytel{\aaref@jnl{S\&T}}             
\def\ssr{\aaref@jnl{Space~Sci.~Rev.}}     
\def\zap{\aaref@jnl{ZAp}}                 
\def\nat{\aaref@jnl{Nature}}              
\def\aplett{\aaref@jnl{Astrophys.~Lett.}} 
\def\apspr{\aaref@jnl{Astrophys.~Space~Phys.~Res.}} 
\def\physrep{\aaref@jnl{Phys.~Rep.}}      
\def\physscr{\aaref@jnl{Phys.~Scr}}       
\def\commat{\aaref@jnl{Comm.~Math.~Phys.}}              
\def\science{\aaref@jnl{Science}}               
\def\cqg{\aaref@jnl{Classical Quant.~Grav.}}            
\def\jpcs{\aaref@jnl{JPCS}}                                     
\def\ijmpd{\aaref@jnl{Int.~J.~Mod.~Phys.~D}}                    
\def\grg{\aaref@jnl{Gen.~Relat.~Gravit.}}               
\def\rpp{\aaref@jnl{Rep.~Prog.~Phys.}}          
\def\npa{\aaref@jnl{Nucl.~Phys.~A}}        
\def\lrr{\aaref@jnl{Living Rev.~Rel.}}                   
\def\jcap{\aaref@jnl{J.~Cosmology Astropart.~Phys.}}    
\def\rmp{\aaref@jnl{Rev.~Mod.~Phys.}}   
\def\epjc{\aaref@jnl{Eur.~Phys.~J.~C}}
\def\plb{\aaref@jnl{~Phy.~Lett.~B}}
\def\mpla{\aaref@jnl{Mod.~Phy.~Lett.~A}}
\def\arxiv{\aaref@jnl{arxiv.org}}

\allowdisplaybreaks[1]

\begin{document}

\color{black}

\title{Late-time cosmology in $f(Q, L_m)$ gravity: Analytical solutions and observational fits}

\author{Yerlan Myrzakulov\orcidlink{0000-0003-0160-0422}}\email[Email: ]{ymyrzakulov@gmail.com} 
\affiliation{Department of General \& Theoretical Physics, L.N. Gumilyov Eurasian National University, Astana, 010008, Kazakhstan.}
\affiliation{Ratbay Myrzakulov Eurasian International Center for Theoretical Physics, Astana, 010009, Kazakhstan.}

\author{O. Donmez\orcidlink{0000-0001-9017-2452}}
\email[Email: ]{orhan.donmez@aum.edu.kw \textcolor{black}{(corresponding author)}}
\affiliation{College of Engineering and Technology, American University of the Middle East, Egaila 54200, Kuwait.}

\author{M. Koussour\orcidlink{0000-0002-4188-0572}}
\email[Email: ]{pr.mouhssine@gmail.com}
\affiliation{Department of Physics, University of Hassan II Casablanca, Morocco.} 

\author{D. Alizhanov}
\email[Email: ]{dilmurod0413@gmail.com}
\affiliation{Department of Physics, Namangan State University
Boburshokh Str. 161, Namangan 160107, Uzbekistan.}

\author{S. Bekchanov}
\email[Email: ]{shukurla@urdu.uz}
\affiliation{Computer science department, Urgench State University, Kh. Alimjan str. 14, Urgench 221100, Uzbekistan.}

\affiliation{Institute of Fundamental and Applied Research, National Research University TIIAME, Kori Niyoziy 39, Tashkent 100000, Uzbekistan.}

\author{J. Rayimbaev\orcidlink{0000-0001-9293-1838}}
\email[Email: ]{javlon@astrin.uz}
\affiliation{Institute of Fundamental and Applied Research, National Research University TIIAME, Kori Niyoziy 39, Tashkent 100000, Uzbekistan.}
\affiliation{University of Tashkent for Applied Sciences, Str. Gavhar 1, Tashkent 100149, Uzbekistan.}
\affiliation{Shahrisabz State Pedagogical Institute, Shahrisabz Str. 10, Shahrisabz 181301, Uzbekistan.}
\affiliation{Tashkent State Technical University, Tashkent 100095, Uzbekistan.}

\begin{abstract}
In this study, we examined the late-time cosmic expansion of the universe within the framework of $f(Q, L_m)$ gravity, where $Q$ denotes the non-metricity and $L_{m}$ represents the matter Lagrangian. We analyzed a linear $f(Q, L_m)$ model of the form $f(Q, L_m) = -\alpha Q + 2 L_{m} + \beta$. Using MCMC methods, we constrained the model parameters $H_0$, $\alpha$, and $\beta$ with various datasets, including $H(z)$, Pantheon+SH0ES, and BAO data. For the $H(z)$ dataset, we found $H_0 = 67.90 \pm 0.66$, $\alpha = 0.1072_{-0.0069}^{+0.0054}$, and $\beta = -1988.2 \pm 1.0$. For the Pantheon+SH0ES dataset, $H_0 = 70.05 \pm 0.68$, $\alpha = 0.0916_{-0.0033}^{+0.0028}$, and $\beta = -1988.3 \pm 1.0$. For the BAO dataset, $H_0 = 68.1 \pm 1.0$, $\alpha = 0.1029_{-0.0052}^{+0.0041}$, and $\beta = -1988.24 \pm 0.99$. Moreover, the energy density remains positive and approaches zero in the distant future, and the deceleration parameter indicates a transition from deceleration to acceleration, with transition redshifts of $z_t = 0.60$, $z_t = 0.78$, and $z_t = 0.66$ for the respective datasets. These findings align with previous observational studies and contribute to our understanding of the universe's expansion dynamics.

\end{abstract}

\maketitle

\tableofcontents

\section{Introduction}\label{sec1}

The present findings from Type Ia supernovae (SNeIa) \cite{Riess/1998,Riess/2004,Perlmutter/1999}, along with studies from the Large Scale Structure (LSS) \cite{T.Koivisto,S.F.}, Wilkinson Microwave Anisotropy Probe (WMAP) \cite{Spergel}, Cosmic Microwave Background Radiation (CMBR)  \cite{R.R.,Z.Y.}, and Baryonic Acoustic Oscillations (BAO) \cite{D.J.,W.J.}, indicate that the universe is currently undergoing an accelerated expansion phase. To address this basic question, dark energy (DE) models find substantial support in the standard model of cosmology. According to references \cite{Zlatev/1999,Weinberg/1989}, the cosmological constant $\Lambda$—which is related to vacuum quantum energy—is the most widely used explanation for DE. Despite $\Lambda$ fitting observational data very well, it faces two main problems: the coincidence problem and the cosmological constant problem \cite{Copeland/2006,Padmanabhan/2003,Steinhardt/1999}. The value predicted by particle physics differs by nearly 120 orders of magnitude from the value needed to match cosmological observations. 

Taking into account the possibility that Einstein's general relativity (GR) may fail at large cosmic scales and require a more universal action to describe the gravitational field is another intriguing way to explain recent discoveries of the universe's cosmic expansion. This perspective opens the door to a variety of alternative theories of gravity that go beyond the standard model. These theories often introduce modifications to the Einstein-Hilbert action or incorporate new dynamical fields that can influence the behavior of gravity on cosmological scales. For example, in $f(R)$ gravity theories, the standard action is replaced by a generic function $f(R)$, where $R$ denotes the Ricci scalar. These theories, as introduced in \cite{Buchdahl/1970}, aim to explain the observed accelerated expansion of the universe without the need for dark energy. The acceleration of the universe during both the early and late stages can be described using $f(R)$ gravity \cite{Dunsby/2010,Carroll/2004}. The $f(R)$ gravity theory was extended in \cite{Bertolami/2007}, where a generic function $f(R)$ and the matter Lagrangian density $L_m$ are explicitly coupled. This matter-geometry coupling forms an additional force orthogonal to the four-velocity vector, which causes the massive particles to move non-geodesically. This model was further extended to include arbitrary couplings in both the matter and geometry sectors \cite{Harko/2008}. Numerous studies have been conducted in \cite{Harko/2010_1,Harko/2010_2,Nesseris/2009,Faraoni/2007,Faraoni/2009} to examine the cosmological and astrophysical consequences of non-minimal matter-geometry couplings. Lately, Harko and Lobo \cite{Harko/2010_3} presented a more advanced generalization of matter-curvature coupling theories, known as $f(R, L_m)$ gravity theory, where $f(R, L_m)$ is an arbitrary function of the matter Lagrangian density $L_m$ and the Ricci scalar $R$. The equivalency principle, which is severely constrained by solar system testing, can be explicitly violated in the $f(R, L_m)$ gravity models \cite{Bertolami}. Recently, Wang and Liao investigated energy conditions within the framework of $f(R, L_m)$ gravity \cite{Wang/2012}. In addition, Goncalves and Moraes analyzed cosmological aspects arising from the non-minimal coupling between matter and geometry, specifically within the context of $f(R, L_m)$ gravity \cite{Goncalves/2023}. By combining observational constraints, Myrzakulova et al. \cite{Myrzakulova/2024} investigated the dark energy issues using $f(R, L_m)$ cosmological models. In a separate study, Myrzakulov et al. \cite{Myrzakulov/2024} investigated the linear redshift parametrization of the deceleration parameter within the framework of $f(R, L_m)$ gravity.

Furthermore, Weyl developed an extension of Riemannian geometry (on which $f(R)$ gravity is based), introducing the first unified theory of gravity and electromagnetism \cite{Weyl/1918}. In this theory, the non-metricity of spacetime gives rise to the electromagnetic field. Consequently, the symmetric teleparallel representation emerges as another generalization of GR. The above generalization is also applicable to non-metric gravity, or $f(Q)$ gravity \cite{Jimenez/2018,Jimenez/2020}, in which the geometric variable $Q$ denotes the non-metricity. This variable characterizes the nature of gravitational interactions and geometrically describes the change in the length of a vector during parallel transport. In the context of $f(Q)$ gravity, Khyllep et al. \cite{Khyllep/2021} looked into cosmic solutions and the growth index of matter perturbations. Research on the cosmology of $f(Q)$ theory, incorporating observational constraints, has shown that the universe's accelerated expansion can be explained without invoking exotic dark energy or additional fields  \cite{MK1, MK2, MK3, MK4, MK5, MK6, MK7, MK8}. Recently, Yixin et al. \cite{Xu/2019,Xu/2020} introduced a new extension of $f(Q)$ gravity, known as $f(Q, T)$ theory, where the non-metricity $Q$ is non-minimally coupled to the trace of the energy-momentum tensor $T$. First, Xu et al. \cite{Xu/2019} investigated the cosmological consequences of the $f(Q, T)$ theory. Using observational constraints to inform their research, further investigations \cite{K6, K7} examined the universe's late-time expansion. In addition, several topics have garnered significant interest, including perturbation analyses \cite{Najera}, inflationary scenarios \cite{Shiravand}, baryogenesis mechanisms \cite{Bhattacharjee}, and astrophysical implications \cite{Tayde1, Sneha2, Bourakadi}.

Motivated by $f(R,L_m)$ theory, we investigate a further extension of $f(Q)$ theory by including the Lagrangian of matter into the Lagrangian density, thus introducing the $f(Q, L_m)$ framework \cite{Hazarika/2024}. This framework allows for both non-minimal and minimal couplings between matter and geometry. Through variation of the action with regard to the metric tensor, Hazarika et al. \cite{Hazarika/2024} derived the general system of field equations. Within this theoretical framework, they also looked into the non-conservation of the matter energy-momentum tensor \cite{fRT1}. Moreover, they derived the generalized Friedmann equations and analyzed cosmic evolution with a flat Friedmann-Lema\^{i}tre-Robertson-Walker (FLRW) metric. The authors investigated two distinct gravity models of the function $f(Q, L_m)$, namely: $f(Q, L_m) = -\alpha Q + 2 L_m + \beta$ and $f(Q, L_m) = -\alpha Q + (2 L_m)^2 + \beta$. Previous studies have explored the implications of the coupling between non-metricity $Q$ and the matter Lagrangian $L_m$ on cosmic dynamics. Harko et al. \cite{Harko/2018} explored various cosmological applications by deriving the evolution equations and adopting specific functional forms for $f(Q)$, including power-law and exponential dependencies of the non-minimal couplings. Mandal and Sahoo \cite{Sanjay/2021} examined the equation of state parameter $\omega$ within the framework of non-minimally coupled $f(Q)$ gravity. Further, Myrzakulov et al. \cite{fQL1} examined the effects of bulk viscosity on late-time cosmic acceleration within an extended $f(Q, L_m)$ gravity framework. They proposed a linear function $f(Q, L_m) = \alpha Q + \beta L_m$ and derived exact solutions under the assumption of non-relativistic matter domination. Here is the corrected version of the paragraph:

The structure of this paper is as follows: In Sec. \ref{sec2}, we describe the basic formalism of $f(Q, L_m)$ gravity. Sec. \ref{sec3} presents the motion equations for the flat FLRW metric. In Sec. \ref{sec4}, we examine a cosmological $f(Q, L_m)$ model and subsequently obtain the expression for both the Hubble parameter and the deceleration parameter in a universe dominated by matter. In Sec. \ref{sec5}, we use the $H(z)$, Pantheon+SH0ES Sample, and BAO datasets to calculate the best fit of the model parameters. We also examine the behavior of the energy density and deceleration parameter for model parameter values that are constrained by observational data.

\section{$f(Q,L_{m})$ gravity theory}\label{sec2}

Weyl–Cartan geometry is a generalization of Riemannian geometry that includes both torsion and non-metricity. In this framework, the affine connection $Y^\alpha_{\;\;\mu\nu}$ can be decomposed into three distinct components: the symmetric Levi-Civita connection $\Gamma^\alpha_{\;\;\mu\nu}$, the contortion tensor $K^\alpha_{\;\;\mu\nu}$, and the disformation tensor $L^\alpha_{\;\;\mu\nu}$. Hence, it can be expressed as follows \cite{Xu/2019}:
\begin{equation}
Y^\alpha_{\;\;\mu\nu}=\Gamma^\alpha_{\;\;\mu\nu}+K^\alpha_{\;\;\mu\nu}+L^\alpha_{\;\;\mu\nu}.
\end{equation}

The Levi-Civita connection $\Gamma^\alpha_{\;\;\mu\nu}$ characterizes curvature and parallel transport in a torsion-free, metric-compatible space. It is determined entirely by the metric $g_{\mu\nu}$ and its first derivatives, encapsulating the conventional gravitational effects in GR,
\begin{equation}
    \Gamma^\alpha_{\;\;\mu\nu}=\frac12 g^{\alpha\lambda}(\partial_\mu g_{\lambda \nu}+\partial_\nu g_{\lambda \mu} - \partial_\lambda g_{\mu\nu}).
\end{equation}

The contortion tensor $K^\alpha_{\;\;\mu\nu}$ introduces the torsion tensor $T^\alpha_{\;\;\mu\nu}$ into the geometry. It measures the deviation from a symmetric connection, reflecting the influence of intrinsic spin or angular momentum of matter on the structure of spacetime. This tensor alters the geodesics, enabling paths that are not solely determined by the metric,
\begin{equation}
    K^\alpha_{\;\;\mu\nu}=\frac{1}{2}(T^\alpha_{\;\;\mu\nu}+ T_{\mu\;\;\nu}^{\;\;\alpha}+T_{\nu\;\;\mu}^{\;\;\alpha}).
\end{equation}

The disformation tensor $L^\alpha_{\;\;\mu\nu}$ represents the non-metricity of the connection, indicating that vector lengths can change under parallel transport. This component is essential in theories where the metric is not preserved during displacement, leading to more general geometric structures that extend beyond the traditional Riemannian framework,
\begin{equation}
    L^\alpha_{\;\;\mu\nu}=\frac{1}{2}(Q^\alpha_{\;\;\mu\nu}-Q^{\;\;\alpha}_{\mu\;\;\nu}-Q^{\;\;\alpha}_{\nu\;\;\mu}).
\end{equation}

For the Weyl–Cartan connection $Y^\alpha_{\;\;\mu\nu}$, the non-metricity tensor $Q_{\alpha\mu\nu}$ is defined as the covariant derivative of the metric tensor. This relationship is given by $Q_{\alpha\mu\nu}=\nabla_\alpha g_{\mu\nu}$, and can be derived as follows:
\begin{equation}
    Q_{\alpha\mu\nu}= \partial_\alpha g_{\mu\nu} - Y^\beta_{\;\;\alpha\mu}g_{\beta\nu}-Y^\beta_{\;\;\alpha\nu}g_{\mu\beta}.
\end{equation}

In addition, we introduce the superpotential $P^\alpha_{\;\;\mu\nu}$, which serves as the conjugate to the non-metricity tensor, defined as follows:
\begin{equation}
    P^\alpha_{\;\;\mu\nu}= -\frac{1}{2}L^\alpha_{\;\;\mu\nu}+\frac{1}{4}(Q^\alpha-\Tilde{Q}^\alpha)g_{\mu\nu}-\frac{1}{4}\delta^\alpha_{\;\;(\mu}Q_{\nu)},
\end{equation}
where $Q^\alpha=Q^{\alpha\;\;\mu}_{\;\;\mu}$ and $\Tilde{Q}^\alpha=Q_{\mu}^{\;\;\alpha\mu}$ denote the non-metricity vectors. By contracting the superpotential tensor with the non-metricity tensor, the non-metricity scalar can be derived:
\begin{equation}
    Q=-Q_{\lambda\mu\nu}P^{\lambda\mu\nu}.
\end{equation}

The gravitational interactions in $f(Q,L_{m})$ gravity are described by the following action \cite{Hazarika/2024}:
\begin{equation}
    S=\int f(Q,L_m) \sqrt{-g} d^4x,\label{Action}
\end{equation}
where $\sqrt{-g}$ denotes the determinant of the metric, and $f(Q, L_m)$ represents an arbitrary function involving the non-metricity scalar $Q$ and the matter Lagrangian $L_m$.

Now, the following field equation can be obtained by varying the action (\ref{Action}) with respect to the metric tensor $g_{\mu\nu}$:
\begin{equation}
\frac{2}{\sqrt{-g}}\nabla_\alpha(f_Q\sqrt{-g}P^\alpha_{\;\;\mu\nu}) +f_Q(P_{\mu\alpha\beta}Q_\nu^{\;\;\alpha\beta}-2Q^{\alpha\beta}_{\;\;\;\mu}P_{\alpha\beta\nu})+\frac{1}{2}f g_{\mu\nu}=\frac{1}{2}f_{L_m}(g_{\mu\nu}L_m-T_{\mu\nu}).\label{field}
\end{equation}

Here, $f_Q \equiv \frac{\partial f}{\partial Q}$, $f_{L_m} \equiv \frac{\partial f}{\partial L_m}$, and $T_{\mu\nu}$ represents the energy-momentum tensor for matter, defined by
\begin{equation}
    T_{\mu\nu}=-\frac{2}{\sqrt{-g}}\frac{\delta(\sqrt{-g}L_m)}{\delta g^{\mu\nu}}=g_{\mu\nu}L_m-2\frac{\partial L_m}{\partial g^{\mu\nu}},
\end{equation}

Again, varying the gravitational action with respect to the connection yields the field equations,
\begin{equation}
    \nabla_\mu\nabla_\nu\Bigl( 4\sqrt{-g}\,f_Q\,P^{\mu\nu}_{\;\;\;\;\alpha}+H_\alpha^{\;\;\mu\nu}\Bigl)=0,
\end{equation}
where $H_\alpha^{\;\;\mu\nu}$ represents the hypermomentum density, defined by $H_\alpha^{\;\;\mu\nu}=\sqrt{-g}f_{L_m}\frac{\delta L_m}{\delta Y^\alpha_{\;\;\mu\nu}}$. Moreover, applying the covariant derivative to the field equation (\ref{field}) enables one to derive,
\begin{equation}
D_\mu\,T^\mu_{\;\;\nu}= \frac{1}{f_{L_m}}\Bigl[ \frac{2}{\sqrt{-g}}\nabla_\alpha\nabla_\mu H_\nu^{\;\;\alpha\mu} + \nabla_\mu\,A^{\mu}_{\;\;\nu} - \nabla_\mu \bigr( \frac{1}{\sqrt{-g}}\nabla_\alpha H_\nu^{\;\;\alpha\mu}\bigr) \Bigr]=B_\nu \neq 0.
\end{equation}

Thus, the matter energy-momentum tensor is not conserved in $f(Q, L_m)$ gravity theory.

\section{Motion equations in $f(Q, L_m)$ gravity} \label{sec3}

For our analysis, we assume the following flat FLRW metric \cite{ryden/2003}, taking into account the spatial isotropy (uniformity in all directions) and homogeneity (uniformity in space) of our universe:
\begin{equation}
\label{FLRW}
    ds^2=-dt^2+a^2(t)(dx^2+dy^2+dz^2).
\end{equation}

Here, $a(t)$ represents the scale factor that quantifies the cosmic expansion at a given time $t$. In the context of the line element (\ref{FLRW}), the non-metricity scalar is $Q = 6 H^2$, where $H = \frac{\dot{a}}{a}$ denotes the Hubble parameter, which signifies the rate of expansion of the universe.

The energy-momentum tensor corresponding to line element (\ref{FLRW}), which describes the universe with perfect fluid-type matter content, is as follows:
\begin{equation}
    T_{\mu\nu}=(\rho+p)u_{\mu}u_{\nu}+pg_{\mu\nu}.
\end{equation}
Here, $\rho$ represents the energy density, $p$ denotes the isotropic pressure, and $u^\mu$ signifies the 4-velocity of the fluid, with components $u^\mu = (1, 0, 0, 0)$. 

The modified Friedmann equations governing the dynamics of the FLRW universe in $f(Q, L_m)$ gravity, assuming the matter content is modeled as a perfect fluid, are formulated as \cite{Hazarika/2024,fQL1}
\begin{eqnarray}
\label{F1}
    && 3H^2 =\frac{1}{4f_Q}\bigr[ f - f_{L_m}(\rho + L_m) \bigl],\\
   && \dot{H} + 3H^2 + \frac{\dot{f_Q}}{f_Q}H=\frac{1}{4f_Q}\bigr[ f + f_{L_m}(p - L_m) \bigl]. \label{F2}
\end{eqnarray}

\section{Cosmological $f(Q, L_m)$ model} \label{sec4}

For our analysis, we consider the following functional form \cite{Hazarika/2024}:
\begin{equation}
    f(Q, L_m) = -\alpha Q +2 L_{m}+\beta,
\end{equation}
where $\alpha$ and $\beta$ are free model parameters. Thus, we find that $f_Q = -\alpha$, and $f_{L_m} = 2$. For this specific $f(Q, L_m)$ model with $L_m = \rho$ \cite{Harko/2015}, the modified Friedmann equations (\ref{F1}) and (\ref{F2}) become
\begin{equation}
3H^2=\frac{\rho}{\alpha}-\frac{\beta}{2\alpha},\label{F11}    
\end{equation}
and
\begin{equation}
2 \Dot{H}+3 H^2=-\frac{p}{\alpha}-\frac{\beta}{2\alpha}. \label{F22}    
\end{equation}

Specifically, for $\alpha =1$ and $\beta = 0$, the usual Friedmann equations of GR can be recovered. From Eqs. (\ref{F11}) and (\ref{F22}), we obtain the following for a matter-dominated universe:
\begin{equation}
\Dot{H}+\frac{3}{2} H^2+\frac{\beta}{4 \alpha}=0. \label{H1}    
\end{equation}

By employing $\frac{1}{H} \frac{d}{dt} = \frac{d}{d \ln(a)}$ (where $a(t) = \frac{1}{1+z}$), we derive the following first-order differential equation:
\begin{equation}
\frac{dH}{d \ln(a)}+\frac{3}{2} H=-\frac{\beta}{4 \alpha}\frac{1}{H}.
\label{H2}    
\end{equation}

By integrating the above equation, we can express the Hubble parameter in terms of redshift $z$ as follows:
\begin{equation}
\label{Hz}
H(z) = \left[ H_{0}^2 (z+1)^3 + \frac{\beta}{6 \alpha} \left\{ (1+z)^3 - 1 \right\} \right]^{\frac{1}{2}}.
\end{equation}

Here, $H_0 = H(z=0)$ represents the current value of the Hubble parameter.

To determine whether the cosmological expansion is accelerating or decelerating, we propose t the deceleration parameter $q$, defined as
\begin{equation}
q = -1 - \frac{\dot{H}}{H^2}.    
\end{equation}

Next, by using Eq. (\ref{Hz}), we obtain
\begin{equation}
\label{qz}
q(z)=-1+\frac{3 (1+z)^3 \left(6 \alpha  H_{0}^2+\beta\right)}{12 \alpha  H_{0}^2 (1+z)^3+2 \beta  z (z (z+3)+3)}.
\end{equation}

From Eq. (\ref{Hz}), it is clear that the form of $H(z)$ reflects a modified cosmological model where the standard term $H_{0}^2 (1+z)^3$ represents the contribution from matter in a flat universe under the standard $\Lambda$CDM model. The additional term $\frac{\beta}{6 \alpha} \left\{ (1+z)^3 - 1 \right\}$ introduces a correction influenced by the parameters $\alpha$ and $\beta$, which could be associated with modifications to gravity, such as $f(Q, L_m)$ gravity. Furthermore, in Eq. (\ref{qz}), the presence of these parameters in both the numerator and denominator indicates their significant role in altering the universe's expansion dynamics compared to the standard model.

\section{Observational constraints} \label{sec5}

In this section, we explore observational interpretations within the current cosmological framework. We use statistical techniques, specifically Markov Chain Monte Carlo (MCMC) methods within the \textit{emcee} Python environment \cite{emcee}, to constrain parameters such as $H_0$, $\alpha$, and $\beta$ using standard Bayesian approaches. In addition, we use the following probability function to optimize the best-fit values for the model parameters:
\begin{equation}
\mathcal{L}\propto \exp(-\chi^2 / 2).
\end{equation}

Here, $\chi^2$ denotes the chi-squared function. We concentrate on three main datasets: Hubble parameter measurements $H(z)$, Pantheon+SH0ES, and BAO. Further, we apply the following priors on the parameters: $60.0 < H_0 < 80.0$, $0 < \alpha < 1$, and $-2500 < \beta < -1500$ to identify the accelerating regime. The MCMC method implementation involves initializing multiple chains with random starting points within the parameter space, which then explore the space based on the likelihood function defined above. The \textit{emcee} package, a widely-used Python library, provides robust tools for parallelizing the MCMC chains and ensuring thorough sampling of the parameter space. In our MCMC study, we employ 100 walkers and 1000
steps to find the fitting outcomes. The $\chi^2$ function employed for different datasets is given below:

\subsection{$H(z)$ dataset}

The Cosmic Chronometers method, a technique for measuring the expansion rate of the universe, employs the differential aging technique to study the oldest and least active galaxies, which are closely separated in redshift, to determine the Hubble rate. The foundation of this method is the definition of the Hubble rate $H = -\frac{1}{1+z}\frac{dz}{dt}$ for an FLRW metric. In this study, we use 31 $H(z)$ data points spanning the redshift range $(0.1, 2)$, sourced from various surveys \cite{Jimenez:2003iv,Simon:2004tf,Stern:2009ep,Moresco:2012jh,Zhang:2012mp,Moresco:2015cya,Moresco:2016mzx,Ratsimbazafy:2017vga}. In order to calculate the best-fit values of the model parameters $H_0$, $\alpha$, and $\beta$, the chi-square function is defined as follows:
\begin{equation}
    \chi_{Hz}^2 = \Delta H^T (C_{Hz}^{-1})\Delta H,
\end{equation}
where $C_{Hz}$ represents the components of the covariance matrix that account for errors in the observed $H(z)$ values, and $\Delta H$ is the vector indicating the difference between the predicted and measured $H(z)$ values for each redshift data point.

\subsection{Pantheon+SH0ES sample}

The Pantheon+SH0ES dataset comprises distance moduli derived from 1701 light curves of 1550 Type Ia supernovae (SNe Ia) gathered from 18 different surveys \cite{Riess:2021jrx,Brout:2022vxf,Brout:2021mpj,Scolnic:2021amr}. These light curves span a redshift range of $0.001 \leq z \leq 2.2613$. Significantly, 77 of these light curves originate from galaxies that also host Cepheids. One of the key advantages of the Pantheon+SH0ES data is its ability to constrain the Hubble constant $H_0$ in conjunction with other model parameters. For the SNe Ia sample, we define the theoretical distance modulus, which is a measure of the distance to an astronomical object, as follows:
\begin{equation}\label{eq:mu}
    \mu(z) = 5\log_{10}\left(\frac{d_L(z)}{1\ \text{Mpc}} \right) + 25.
\end{equation}

Here, $d_L(z)$ represents the luminosity distance, which is the distance to an astronomical object as inferred from its observed brightness, and is expressed as
\begin{equation}\label{eq:d_L}
    d_L(z) = \frac{c(1+z)}{H_0}\int_0^z \frac{dy}{E(y)}.
\end{equation}
where $c$ is the speed of light and $E(z) = \frac{H(z)}{H_0}$ is the dimensionless Hubble parameter. Therefore, the distance residual $\Delta \mu$ is given by $\Delta\mu_i=\mu_i-\mu_{th}(z_i)$. The parameters $H_0$ and $M$ become degenerate when analyzing data from the SNe Ia sample. To address this issue, the SNe Ia distance residuals are adjusted as follows \cite{Perivolaropoulos:2023iqj,Brout:2022vxf}:
\begin{equation}
\Delta\Tilde{\mu}=\begin{cases}
\mu_i-\mu_i^{Ceph}, & \text{if $i$ belongs to Cepheid hosts}\\
\mu_i-\mu_{\text{th}}(z_i), & \text{otherwise}
    		 \end{cases}
\end{equation}
    
In this context, $\mu_i^{Ceph}$ denotes the distance modulus of the Cepheid host for the $i^{th}$ SNe Ia, and $\Delta\tilde{\mu}$ represents the adjusted distance residual. For the Pantheon+SH0ES sample, the $\chi^2$ function is given by
\begin{equation}\label{Eq:ChiSN}
        \chi^2_{SNe}= \Delta\mu^T (C_{\text{stat+sys}}^{-1})\Delta\mu.
\end{equation}

\subsection{BAO dataset}

Lastly, we use BAO to constrain our model \cite{BAO1,BAO2,BAO3,BAO4,BAO5,BAO6}, which are periodic fluctuations in the density of the visible baryonic matter of the universe. To derive the BAO constraints, we employ the acoustic scale $l_{A} = \pi \frac{d_{A}(z_{d})}{r_{s}(z_{*})}$, where $d_A(z)=\int ^z_0\frac{dz'}{H(z')}$ is the angular diameter distance in the comoving coordinates and $r_s$ is the sound horizon determined as $r_s(z_d)=\int_{z_d}^{\infty}\frac{c_s(z')}{H(z)}$ at the drag epoch $z_d$ with sound speed $c_s(z)$. In this case, $D_V(z_{BAO})$ represents the dilation scale, expressed as: $D_V(z)=\left[\frac{d_A(z)^2 cz}{H(z)}\right]^{1/3}$. The BAO constraints are then given by $\frac{d_{A}(z_{d})}{D_{V}(z_{BAO})}$. In this measurement:
\begin{itemize}
    \item The acoustic scale $l_{A}$ quantifies the characteristic scale of BAO imprinted in the cosmic microwave background.
    \item The angular diameter distance $d_A(z)$ is the distance derived from the apparent size of an object of known physical size at redshift $z$.
    \item The sound horizon $r_s$ is the maximum distance that sound waves could travel in the early universe before recombination.
    \item The dilation scale $D_V$ relates to the combined constraints of the angular diameter distance and the Hubble parameter, capturing the overall stretching of scales in the universe's expansion.
\end{itemize}

\subsection{Results}

\begin{table*}[ht]
\begin{ruledtabular}
		    \centering
		    \begin{tabular}{c c c c c}
				Dataset & $H_{0}$ & $\alpha$ & $\beta$\\ 
				\hline
				$H(z)$ & $67.90\pm 0.66$ & $0.1072_{-0.0069}^{+0.0054}$ & $-1988.2\pm 1.0$\\
				Pantheon+SH0ES  & $70.05\pm 0.68$  & $0.0916_{-0.0033}^{+0.0028}$& $-1988.3\pm 1.0$\\
				BAO & $68.1\pm 1.0$ & $0.1029_{-0.0052}^{+0.0041}$ & $-1988.24\pm 0.99$\\
			\end{tabular}
		   \end{ruledtabular}
\caption{The table presents the best-fit values for the model parameters $H_0$, $\alpha$, and $\beta$, as determined from the H(z), Pantheon+SH0ES, and BAO datasets.}
\label{tab}%
\end{table*}

\begin{figure}[h]
\centering
\includegraphics[width=0.8\linewidth]{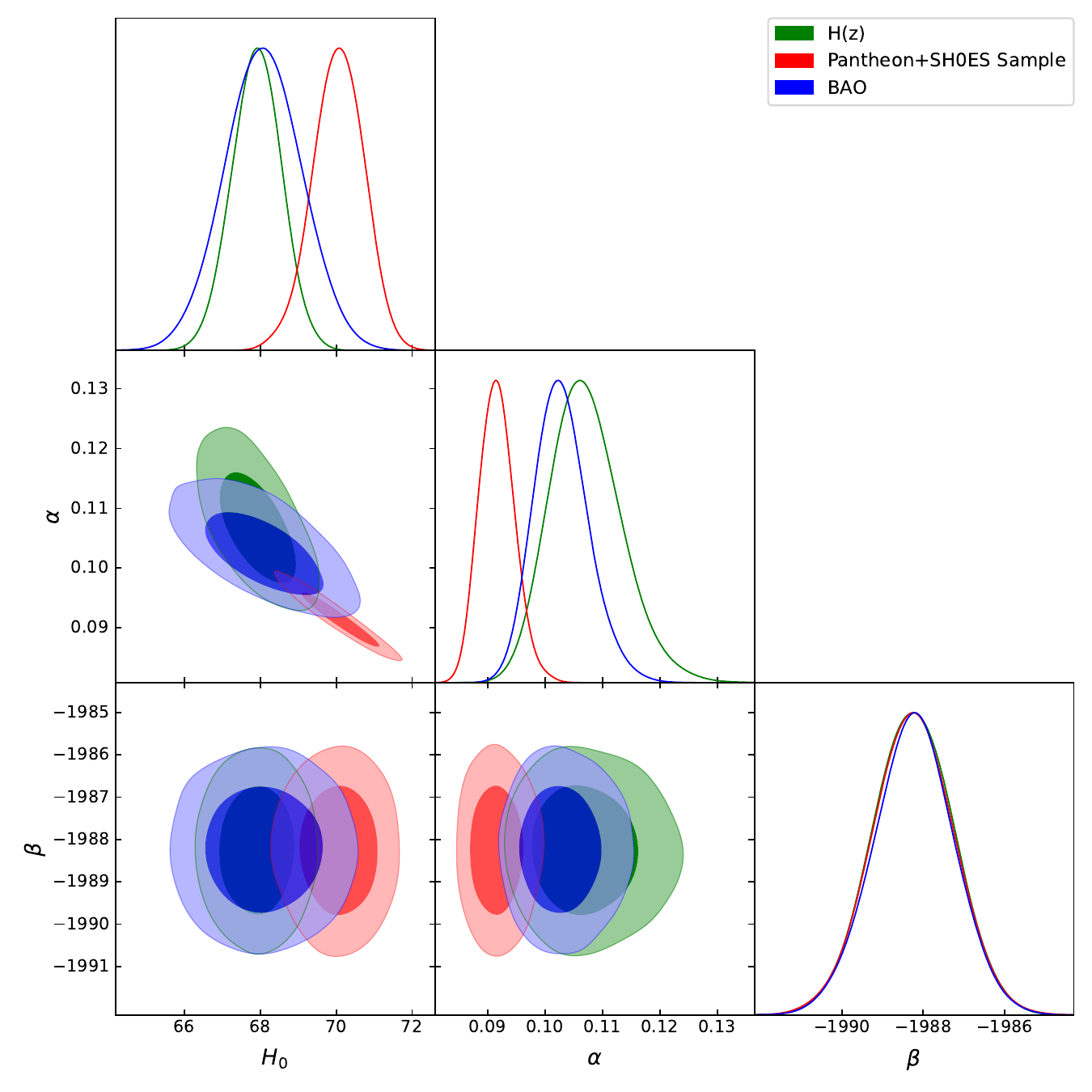}
\caption{The two-dimensional contour plots display the model parameters $H_0$, $\alpha$, and $\beta$, including $1-\sigma$ and $2-\sigma$ error regions. In addition, the plots show the best-fit values for these parameters obtained from the $H(z)$, Pantheon+SH0ES, and BAO datasets.}
\label{F_Com}
\end{figure}

At this stage, we have analyzed various datasets and obtained constrained values for the model parameters $H_0$, $\alpha$ and $\beta$. This includes applying statistical methods to data from $H(z)$, Pantheon+SH0ES, and BAO to refine our understanding of the cosmological model. In addition, we have produced two-dimensional likelihood contours showing $1-\sigma$ and $2-\sigma$ errors, which correspond to 68\% and 95\% confidence levels, respectively, for the $H(z)$, Pantheon+SH0ES, and BAO datasets. These contours are depicted in Fig. \ref{F_Com}. From these figures, it is clear that the likelihood functions for all datasets closely follow a Gaussian distribution. This indicates that the errors associated with the parameter estimates are symmetrically distributed around the best-fit values, with the distribution showing a peak at the most probable values. Initially, we analyzed the $H(z)$ dataset, which consists of 31 data points. For the Hubble constant $H_0$, we obtained a value of $67.90\pm 0.66$ \cite{Planck/2020}. For the parameters $\alpha$ and $\beta$, which are essential for inducing the acceleration phase in modified gravity,  the constrained values are $0.1072_{-0.0069}^{+0.0054}$ and $-1988.2\pm 1.0$, respectively. Subsequently, for the Pantheon+SH0ES dataset, comprising 1048 sample points, we found $H_0=70.05\pm 0.68$, $\alpha=0.0916_{-0.0033}^{+0.0028}$, and $\beta=-1988.3\pm 1.0$. Finally, for BAO datasets, the values are $H_0=68.1\pm 1.0$, $\alpha=0.1029_{-0.0052}^{+0.0041}$, and $\beta=-1988.24\pm 0.99$. The constrained parameter values have significant implications for future cosmological observations and studies. The Hubble constant $H_0$, derived from different datasets, provides a consistency check against other cosmological models and observations, contributing to the ongoing debate on the Hubble tension. The parameters $\alpha$ and $\beta$ provide information about how matter and geometry interact, which may help gravitational theories be improved. Future surveys and observations can leverage these constraints to test the predictions of $f(Q, L_m)$ gravity, potentially leading to new discoveries in cosmology. The summarized constraint values are presented in Tab. \ref{tab}. 

\begin{figure}[h]
\centering
\includegraphics[width=0.6\linewidth]{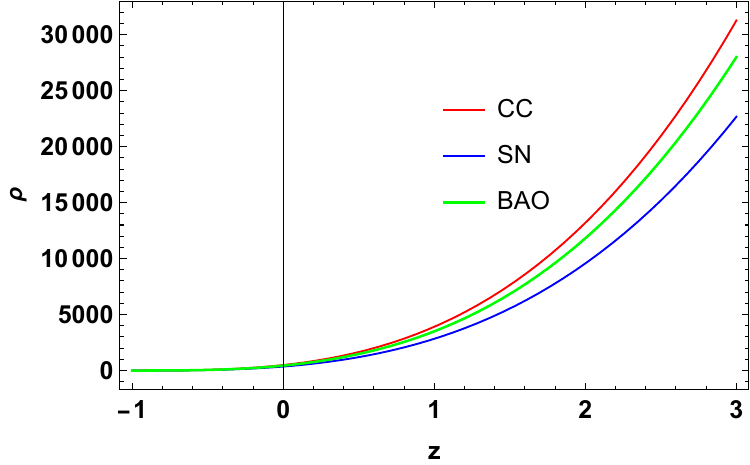}
\caption{The variation of the energy density with respect to redshift $z$ for the best-fit values of the model parameters $H_0$, $\alpha$, and $\beta$, as determined from the $H(z)$, Pantheon+SH0ES, and BAO datasets.}
\label{F_rho}
\end{figure}

\begin{figure}[h]
\centering
\includegraphics[width=0.6\linewidth]{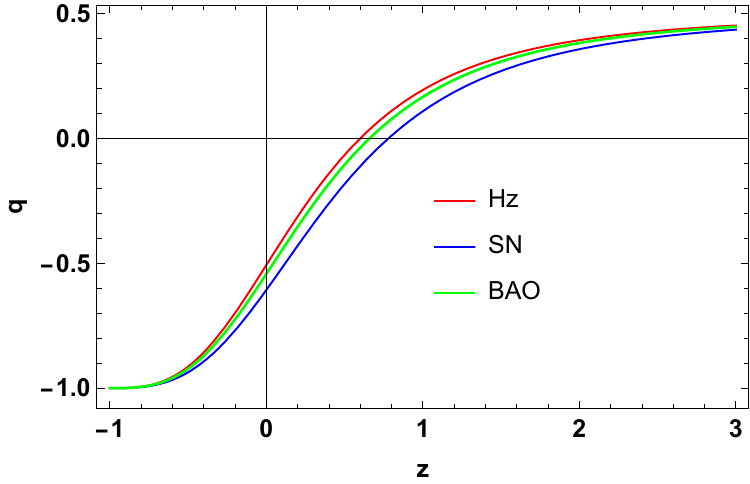}
\caption{The variation of the deceleration parameter with respect to redshift $z$ for the best-fit values of the model parameters $H_0$, $\alpha$, and $\beta$, as determined from the $H(z)$, Pantheon+SH0ES, and BAO datasets.}
\label{F_q}
\end{figure}

Fig. \ref{F_rho} shows that the energy density of the cosmic fluid remains positive and approaches zero $(\rho \to 0)$ in the distant future for all datasets. This behavior is consistent with the predictions of a universe transitioning to a state dominated by dark energy or a similar component, causing an accelerated expansion. The positive energy density throughout indicates that the model remains physically viable over time. As the energy density approaches zero $(\rho \to 0)$, it suggests that the influence of matter and radiation diminishes, leaving a universe where dark energy becomes the dominant factor. This result is crucial for understanding the long-term fate of the universe and provides strong support for the model's predictions. Moreover, the deceleration parameter $q$ plays a crucial role in describing the nature of the universe's expansion, indicating whether it is accelerating ($q < 0$) or decelerating ($q > 0$). From Fig. \ref{F_q}, it is evident that our universe has transitioned from a decelerated phase to an accelerated phase in the recent past. This transition is crucial for understanding the current state of cosmic expansion. The specific transition redshift values, constrained by the $H(z)$, Pantheon+SH0ES, and BAO datasets, are $z_t = 0.60$, $z_t = 0.78$, and $z_t = 0.66$, respectively. Moreover, the present value of the deceleration parameter is found to be $q_0 = -0.50$ for the $H(z)$ datasets, $q_0 = -0.61$ for the Pantheon+SH0ES datasets, and $q_0 = -0.54$ for the BAO datasets. Our findings align with several observational studies \cite{Hernandez,Basilakos,Roman,Jesus,Cunha,Y/2024}.

\section{Concluding remarks}\label{sec6}

In this study, we examined the late-time cosmic expansion of the universe within the framework of $f(R, L_m)$ gravity, where $Q$ denotes the non-metricity and $L_{m}$ represents the matter Lagrangian. We analyzed a linear $f(Q, L_m)$ model of the form $f(Q, L_m) = -\alpha Q + 2 L_{m} + \beta$, with $\alpha$ and $\beta$ as free parameters \cite{Hazarika/2024}. Next, we derived the equations of motion for a flat FLRW universe and obtained an exact solution for the Hubble parameter as a function of redshift within our cosmological $f(Q, L_m)$ model, assuming a matter-dominated universe. 

Using this framework, we applied MCMC methods within the emcee Python environment to constrain the model parameters $H_0$, $\alpha$, and $\beta$. Furthermore, we employed various datasets, including the Hubble parameter measurements $H(z)$, Pantheon+SH0ES, and BAO data. Our results show that the likelihood functions for all datasets closely follow a Gaussian distribution, indicating a reliable estimate of parameter values. The analysis revealed the following results: For the $H(z)$ dataset, the Hubble constant $H_0$ was found to be $67.90 \pm 0.66$, with $\alpha$ and $\beta$ constrained to $0.1072_{-0.0069}^{+0.0054}$ and $-1988.2 \pm 1.0$, respectively. For the Pantheon+SH0ES dataset, the values were $H_0 = 70.05 \pm 0.68$, $\alpha = 0.0916_{-0.0033}^{+0.0028}$, and $\beta = -1988.3 \pm 1.0$. For the BAO dataset, the results were $H_0 = 68.1 \pm 1.0$, $\alpha = 0.1029_{-0.0052}^{+0.0041}$, and $\beta = -1988.24 \pm 0.99$. The summarized constraint values are provided in Tab. \ref{tab}. Therefore, the constraints on $\alpha$ and $\beta$ are consistent across datasets, reinforcing the validity of our model. Further, we examined the behavior of the energy density and deceleration parameter for the constrained values of the model parameters. We observed that the energy density of the cosmic fluid remains positive and approaches zero in the distant future, suggesting a transition to a dark energy-dominated universe. The deceleration parameter confirms a recent transition from deceleration to acceleration in cosmic expansion, with transition redshifts of $z_t = 0.60$, $z_t = 0.78$, and $z_t = 0.66$ for the respective datasets. Cunha and Lima \cite{Cunha} investigated the transition redshift, using SNe Ia data to establish new kinematic constraints. They estimated the transition redshift to be $z_t = 0.61$, a result that aligns closely with the findings of Davis et al. \cite{Davis/2007}, who reported $z_t = 0.60$. Our findings have broader implications for the field of cosmology, particularly in the context of modified gravity theories. The consistency of the parameter constraints across different datasets underscores the viability of $f(Q, L_m)$ gravity as an alternative to the standard $\Lambda$CDM model. This work not only advances our understanding of late-time cosmic acceleration but also opens avenues for exploring the interplay between geometry and matter in shaping the universe's evolution. As observational capabilities continue to improve, the insights gained from $f(Q, L_m)$ gravity models will be crucial in addressing fundamental questions about the nature of gravity and the universe.

While our study provides robust constraints on the $f(Q, L_m)$ model parameters, several limitations should be noted. Our analysis is based on a specific functional form of $f(Q, L_m) = -\alpha Q + 2 L_m + \beta$, and exploring other functional forms, such as $f(Q, L_m) = -\alpha Q + (2 L_m)^2 + \beta$ \cite{Hazarika/2024}, could yield different insights. In addition, the assumption of a perfect fluid may overlook complexities such as anisotropies and inhomogeneities. Future research could extend our work by incorporating these factors, as well as by using more diverse and higher-resolution datasets to refine the constraints on the model parameters.

\section*{Acknowledgment}
This research was funded by the Science Committee of the Ministry of Science and Higher Education of the Republic of Kazakhstan (Grant No. AP22682760).

\section*{Data Availability Statement}
This article does not introduce any new data.

\end{document}